\documentclass[pra,showpacs,amsmath,amssymb,amsfonts,superscriptaddress,lengthcheck,longbibliography]{revtex4-2}

\usepackage[letterpaper,top=2cm,bottom=2cm,left=3cm,right=3cm,marginparwidth=1.75cm]{geometry}
\usepackage[dvipsnames]{xcolor}
\usepackage{tcolorbox}
\tcbuselibrary{breakable}
\usepackage{graphicx, color, graphpap}
\usepackage{enumitem}
\usepackage{amssymb}
\usepackage{amsthm}
\usepackage{dsfont}
\usepackage{mathtools}
\usepackage{multirow}
\usepackage[colorlinks=true,citecolor=magenta,linkcolor=blue]{hyperref}
\usepackage[T1]{fontenc}
\usepackage{bbm}
\usepackage{thmtools,thm-restate}
\usepackage{verbatim}
\usepackage{mathtools}
\usepackage{titlesec}
\usepackage{amsmath}
\usepackage[normalem]{ulem}
\usepackage{circuitikz}
\usepackage{braket}
\usepackage[caption=false]{subfig} 
\usepackage{anyfontsize}
\usepackage{verbatim}

\usepackage{xcolor}

\begin{document}

\title{Solving the compute crisis with physics-based ASICs}
\author{Maxwell Aifer}
\affiliation{Normal Computing Corporation, New York, New York, USA}
\author{Zach Belateche}
\affiliation{Normal Computing Corporation, New York, New York, USA}
\author{Suraj Bramhavar}
\affiliation{Advanced Research + Invention Agency (ARIA)}
\author{Kerem Y. Camsari}
\affiliation{Department of Electrical and Computer Engineering,
University of California, Santa Barbara, Santa Barbara, CA 93106, USA}
\author{Patrick~J.~Coles}
\affiliation{Normal Computing Corporation, New York, New York, USA}
\author{Gavin~Crooks}
\affiliation{Normal Computing Corporation, New York, New York, USA}
\author{Douglas J. Durian}
\affiliation{Department of Physics and Astronomy, University of Pennsylvania, Philadelphia, PA 19104, USA}
\affiliation{Department of Mechanical Engineering and Applied Mechanics,
University of Pennsylvania, Philadelphia, PA 19104, USA}
\author{Andrea J. Liu}
\affiliation{Department of Physics and Astronomy, University of Pennsylvania, Philadelphia, PA 19104}
\affiliation{Santa Fe Institute, 1399 Hyde Park Road, Santa Fe, NM 87501, USA}
\author{Anastasia Marchenkova}
\affiliation{Normal Computing Corporation, New York, New York, USA}
\author{Antonio~J.~Martinez}
\affiliation{Normal Computing Corporation, New York, New York, USA}
\author{Peter L. McMahon}
\affiliation{School of Applied and Engineering Physics, Cornell University, Ithaca, NY 14853, USA}
\affiliation{Kavli Institute at Cornell for Nanoscale Science, Cornell University, Ithaca, NY 14853, USA}
\author{Faris Sbahi}
\affiliation{Normal Computing Corporation, New York, New York, USA}
\author{Benjamin Weiner}
\affiliation{Advanced Research Projects Agency - Energy (ARPA-E), Washington, D.C. 20024, USA}
\author{Logan G. Wright}
\affiliation{Department of Applied Physics, Yale University, New Haven, CT 06520, USA}

\begin{abstract}
Escalating artificial intelligence (AI) demands expose a critical ``compute crisis'' characterized by unsustainable energy consumption, prohibitive training costs, and the approaching limits of conventional CMOS scaling. Physics-based Application-Specific Integrated Circuits (ASICs) present a transformative paradigm by directly harnessing intrinsic physical dynamics for computation rather than expending resources to enforce idealized digital abstractions. By relaxing the constraints needed for traditional ASICs, like enforced statelessness, unidirectionality, determinism, and synchronization, these devices aim to operate as exact realizations of physical processes, offering substantial gains in energy efficiency and computational throughput. This approach enables novel co-design strategies, aligning algorithmic requirements with the inherent computational primitives of physical systems. Physics-based ASICs could accelerate critical AI applications like diffusion models, sampling, optimization, and neural network inference as well as traditional computational workloads like scientific simulation of materials and molecules. Ultimately, this vision points towards a future of heterogeneous, highly-specialized computing platforms capable of overcoming current scaling bottlenecks and unlocking new frontiers in computational power and efficiency.\footnote{All claims expressed in this article are solely those of the authors and do not necessarily represent those of their affiliated organizations.}
\end{abstract}

\maketitle

\section{Introduction: The Compute Crisis}
Over the past decade, the rapid expansion of artificial intelligence (AI) applications has significantly increased demands on computing infrastructure, revealing critical limitations in the underlying hardware paradigm. The infrastructure powering AI models was never designed with today’s scale, complexity, or energy demands in mind. As a result, the current computing stack results in substantial under-utilization of the raw physical computing power inherent in current hardware systems. 

\begin{figure}[t]
    \centering   \includegraphics[width=0.49\textwidth]{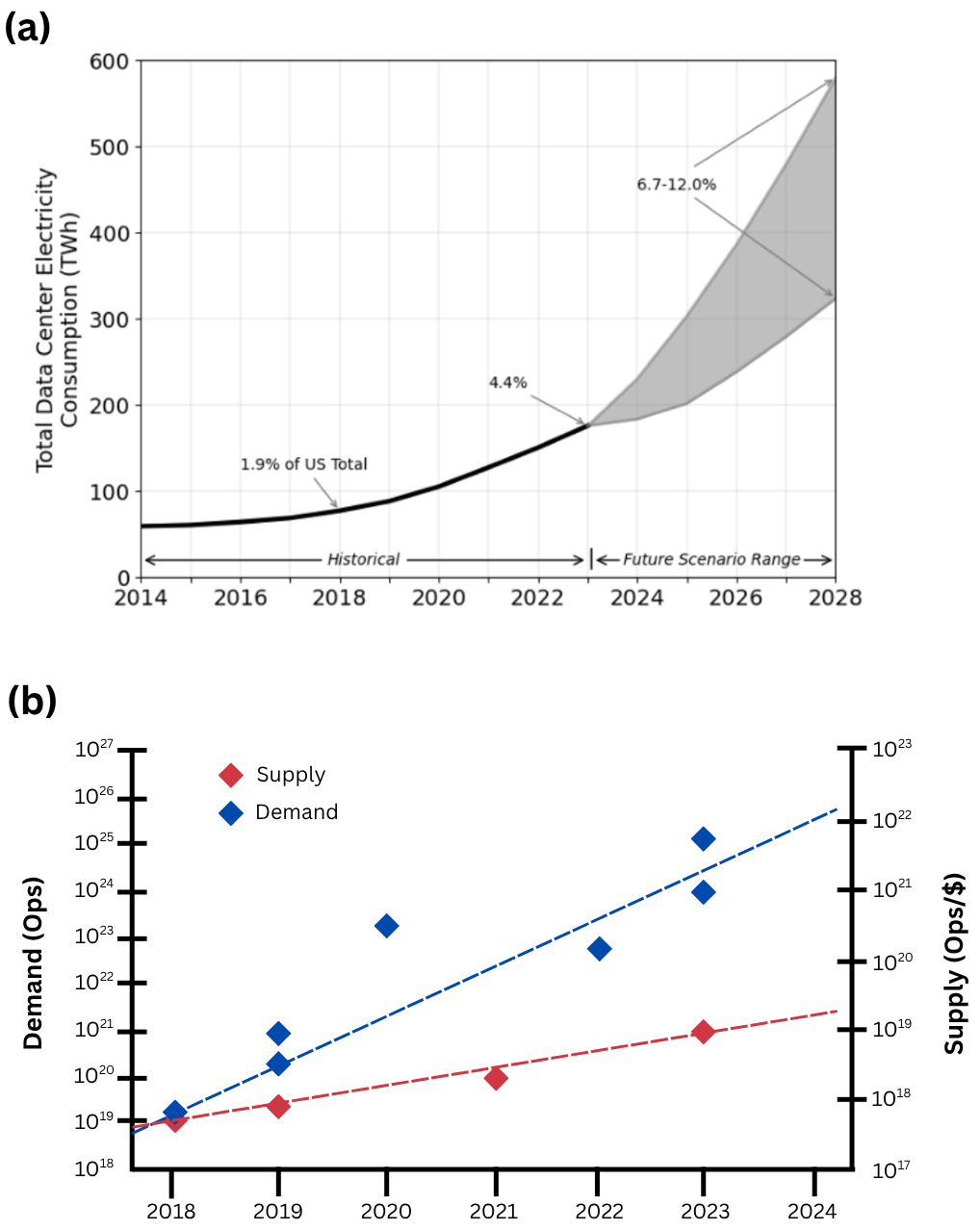}   
    \caption{\textbf{Projected energy consumption and supply/demand for computing power.} While the compute crisis has many facets, two key aspects are (a) the rising energy consumption of compute and (b) the growing gap between the supply and demand for compute (shown here for the case of AI model training). Both of these issues have largely been driven by the AI revolution over the past several years. Panels (a) and (b) were respectively adapted from Refs.~\cite{lbnl2024} and \cite{epoch2025}.}
    \label{fig:crisis}
\end{figure}

Conventional scaling is reaching its limits across multiple axes: 
\begin{enumerate}
    \item Energy demands from AI are escalating unsustainably, as shown in Fig.~\ref{fig:crisis}(a). Data centers, which are central to AI operations, consumed approximately 200 terawatt-hours (TWh) of electricity in 2023. Projections indicate this could rise to 260 TWh by 2026, accounting for about 6\% of total U.S. electricity demand \cite{GartnerAIDataCenterShortage}.
    \item Compute cost is rising steeply, centralizing access. The development of frontier AI models has seen training costs escalate dramatically, with estimates suggesting that the largest training runs will cost more than \$1B dollars by 2027 \cite{Owen2024AIModelCost}. This is naturally connected to the gap between supply and demand shown in Fig.~\ref{fig:crisis}(b).
    \item  As transistor dimensions shrink to the nanometer scale, the long-standing scaling laws -- Moore's Law and Dennard's Law -- are reaching their limits. Miniturization effects such as stochasticity, leakage currents, and variability make reliable operation difficult at these scales. We can no longer proportionally reduce the threshold voltage as we reduce feature size, leading to higher power densities that, in turn, result in heating that limits clock speeds and runtime.
\end{enumerate}

These constraints not only hinder performance but also reveal a deeper inefficiency: today’s general-purpose architectures underutilize the physical potential of the hardware itself. The abstraction layers designed to manage complexity now act as bottlenecks, especially in energy efficiency and computational throughput. Without a shift in the computing paradigm, we risk stagnating innovation, escalating energy costs, and the concentration of AI capabilities within a limited number of large corporations and governmental entities.

Physics-based Application-Specific Integrated Circuits (ASICs) offer a transformative approach by leveraging, rather than suppressing, physical phenomena for computation. By aligning hardware design with the intrinsic properties of physical systems, these ASICs can enhance efficiency, reduce energy consumption, and democratize access to AI and computational resources.

\section{What Are Physics-Based ASICs?}

\subsection{Motivation}

If we want to improve computation (e.g., to cost less energy or require less time), we can design more efficient algorithms for an idealized, general-purpose hardware, create faster or more efficient hardware (whether general-purpose or not), or jointly (co)design algorithms and hardware together, aiming to maximize the useful computation obtained. While there are many exceptions within the contemporary research landscape of computer science and engineering, explicit efforts to improve computation over the last half-century or so have concentrated primarily on the first two routes, a strategy of general-purpose computing hardware and highly abstracted software development that has enabled ever-expanding software applications and our modern digital economy. But more specialized hardware, such as GPUs, have nonetheless emerged as a key driving force for computing's more recent advancement, and the implicit algorithmic preferences of hardware have long served as a guiding force for the success of algorithms. Is it a coincidence that the most popular algorithms for machine learning happen to involve primarily matrix-matrix multiplications, an operation GPUs are especially efficient at? Of course not: the exceptional match between software and hardware these algorithms achieve allows them to scale well, achieving better results than algorithms that less effectively utilized GPUs. This general tendency, in which the communal optimization of algorithms is implicitly guided by the idiosyncracies of available hardware, is known as the ``the hardware lottery~\cite{hooker2021hardware}''. The hardware lottery's prominence shows that software-hardware codesign is inevitable, purposeful or not. 

The idea of Physics-based ASICs is essentially to make this largely unintentional trend instead fully intentional and principled: it aims to deliberately codesign algorithms and hardware down to the lowest possible physical level of available, scalable hardware infrastructure. Similar to the way in which the dense matrix-matrix multiplications of Transformers have cleverly played to the preferences of GPUs, could we analogously design algorithms and electronic chips that exploit even deeper preferences within the physics of silicon electronic circuits (and in turn, unlock even greater scalability)? Of course, this is not a free lunch: it will require development of new algorithms and hardware that, unlike those designed by most modern computer scientists, will need to consider intimate details of each other. But on the other hand, this path may allow us to utilize modern computing hardware vastly more efficiently than it is used today. How much more efficiently? This is difficult to say, but we can get a hint by considering a related question, namely how abstractions affect the cost of \textit{digitally simulating} circuitry. For example, the physical device that performs a simple CMOS NOT gate realizes a single binary operation per clock cycle when abstracted as a binary logic gate, but if we instead model the transient (and analog) dynamics of the circuit that comprises it, typical numerical methods (e.g., used in SPICE) may require millions of floating point operations. If we modeled each transistor in microscopic detail (as is often done during the design phase), we would necessarily solve 3+1-dimensional systems of partial differential equations, requiring billions or even trillions of floating point operations (still just for a single clock period). Clearly, the physical level at which we abstract a physical system can affect how many digital logic gate operations it is equivalent to. However, this is merely part of the challenge: just because a simulation of a physical system at some level of abstraction is expensive does not necessarily imply we can use that same physical system and that abstraction to then perform other calculations of interest. This is the core challenge of Physics-based ASICs: to design abstractions, algorithms, and hardware architectures that, by better respecting the underlying hardware physics, allow us to usefully and more fully exploit the available physical computation of today's highly scalable electronic circuits.

\subsection{Definition}

Loosely speaking, physics-based ASICs are ASICs that rely on the natural physical dynamics of a system to carry out non-trivial operations on data. This definition is somewhat ambiguous; as all circuits are governed by the laws of physics, all computations in some sense are accomplished via the natural evolution of the computing system.

\begin{figure}[t]
    \centering
    \includegraphics[width=0.49\textwidth]{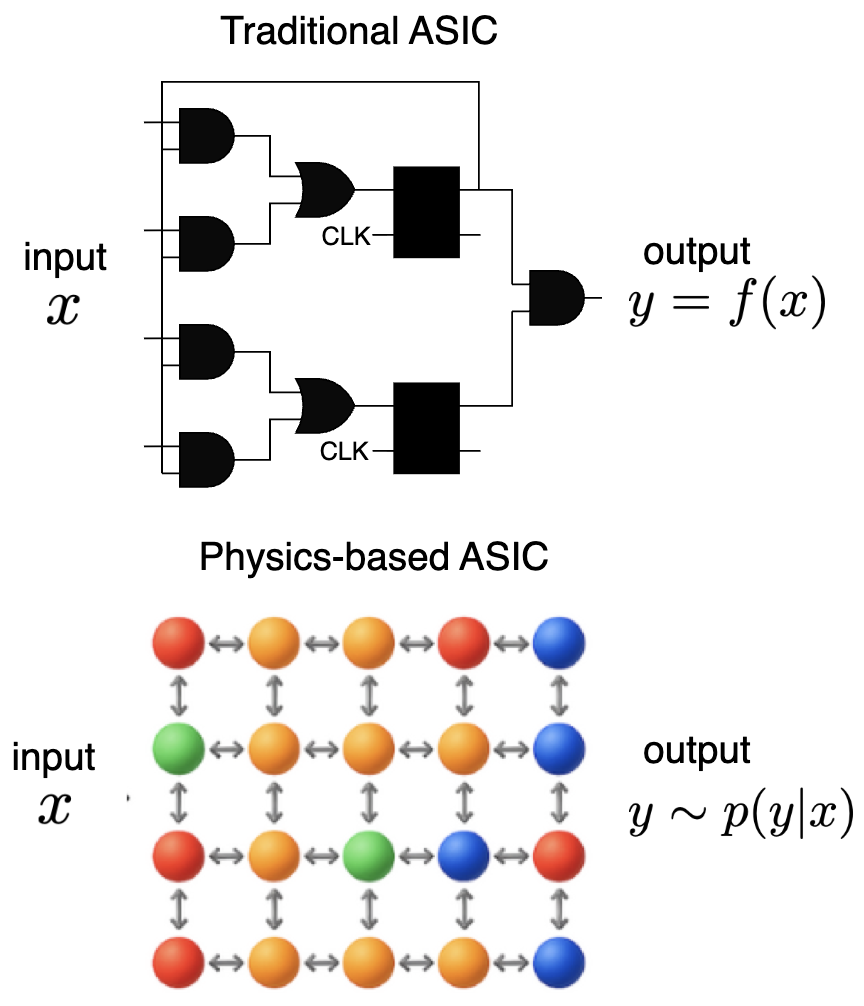}
    \caption{\textbf{Traditional ASICs vs. Physics-based ASICs.} As illustrated, traditional ASICs exhibit a separation of memory and computation, with computational components assumed to be stateless. Individual logic gates carry information in a single direction, having dedicated input and output terminals, and it is necessary to make a connection from output to input to create feedback loops. Physics-based ASICs may have stateful computational components as well as bidirectional information flow across couplings. }
    \label{fig:comparison}
\end{figure}

However, conventional ASIC design intentionally suppresses or abstracts away certain physical effects in order to realize idealized, symbolic models of computation. In doing so, it relies on a set of approximations that allow complex systems to be built from simple, idealized components. Among the most important are:

\begin{enumerate}

\item Statelessness: In conventional ASICs there is usually a clear separation between memory and computation, which are handled by separate components in different locations. Components that are not responsible for storing information are assumed to behave as if their outputs depend only on their current inputs, not on prior history. For example, a NOT gate should invert the present value of its input, regardless of past values.

\item Unidirectionality: Primitive components of conventional ASICs are designed to propagate information in a single direction; they have designated input and output terminals. For example, a NOT gate should respond to changes at its input, but its output should not affect its input. Because of this, creating feedback loops in conventional ASICs requires explicitly connecting the output of some block to its input.

\item Determinism: Given identical inputs and initial conditions, the circuit is expected to produce the same outputs every time.

\item Synchronization: Usually, signals in different parts of a conventional ASIC are synchronized with each other according to a centralized clock.
\end{enumerate}

These properties are not physically realizable in an exact sense: real components exhibit memory effects, feedback, noise, and thermal fluctuations. Enforcing these ideal behaviors incurs energy, latency, or complexity costs, which grow as the approximation becomes more precise.

Physics-based ASICs are instead designed to function without relying on these properties (or at least without some of them). In contrast to conventional ASICs, these devices are designed to take advantage of (or at least tolerate) statefulness, bidirectionality, non-determinism, and asynchronization, as illustrated in Fig.~\ref{fig:comparison}. Therefore, a computation on a physics-based ASIC is not an approximation to a non-physical process but rather the realization of a physical process.

Due to the lack of the simplifying assumptions present in conventional ASICs, the behavior of physics-based ASICs is often more physically complex and more difficult to analyze. However, there is also a greater range of possibilities for the operations performed by circuit components in physics-based ASICs. As a result, physics-based ASICs are often able to accomplish significantly more computation with a smaller number of components. For example, a scalar multiplication may require tens to hundreds of transistors in a conventional ASIC, but only require a handful of components on a physics-based ASIC.

\subsection{Platforms}

\begin{figure}[t]
    \centering
    \includegraphics[width=0.49\textwidth]{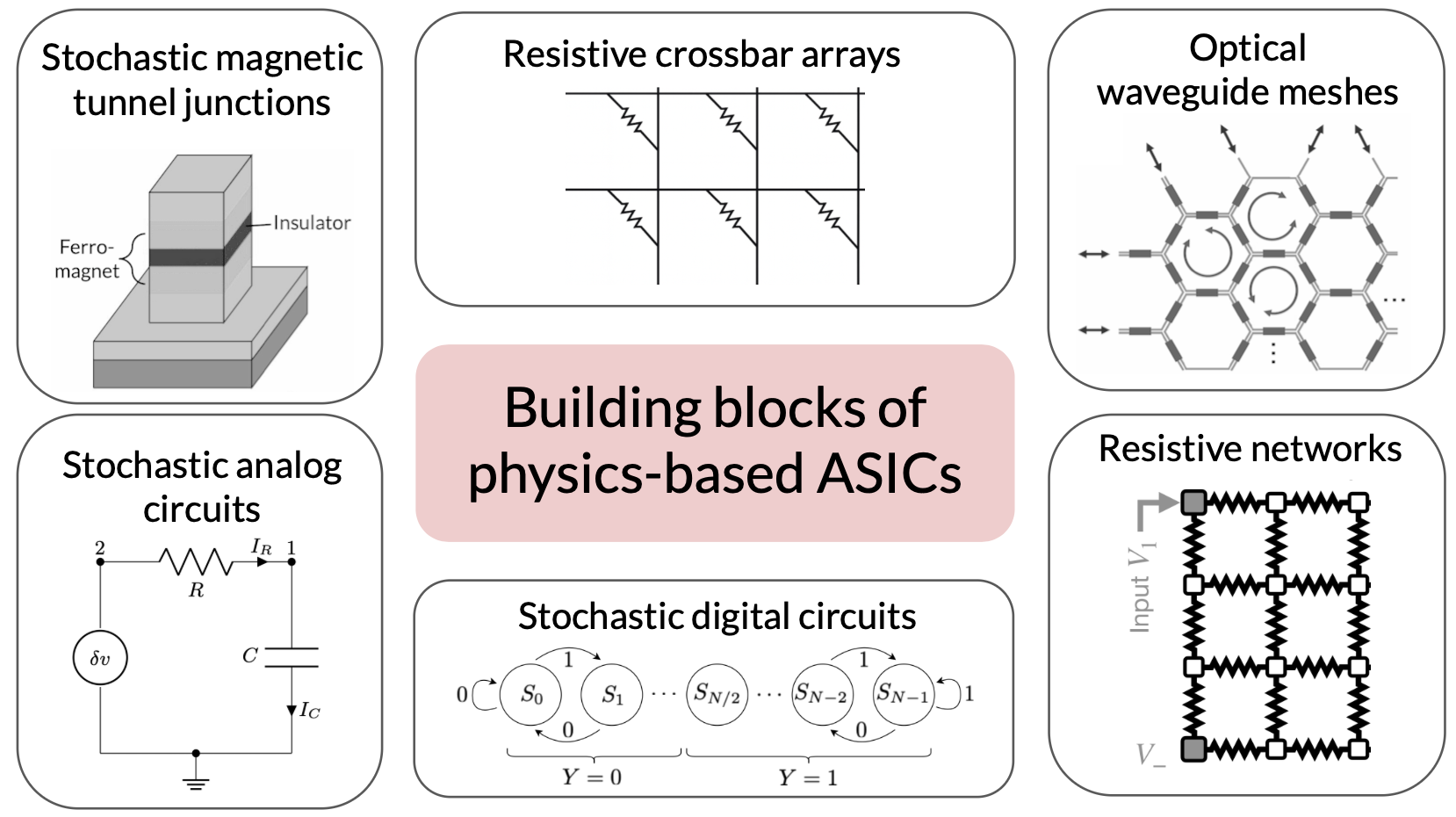}
    \caption{\textbf{Common building blocks of physics-based ASICs.} While not an exhaustive list, several basic physical structures are shown that can be used as building blocks for physics-based ASICs. For each of these components, the physical laws describing the device can be mapped to some computational primitive operation. Panels on the top right, bottom right, and bottom left were respectively adapted from Ref.~\cite{bogaerts2020programmable}, Ref.~\cite{dillavou2024machine}, and Ref.~\cite{coles2023thermodynamic_published}.}
    \label{fig:buildingblocks}
\end{figure}

Many existing unconventional computing paradigms can be viewed as examples of physics-based ASICs. While there is a great diversity among these different approaches, physics-based ASICs are distinguished from other physics-based platforms (e.g., computing with soap bubbles~\cite{aaronson2005guest}) by their scalability. Scalability and manufacturability are key ingredients in this exciting new field. Let us now give examples of these scalable platforms, some of which are illustrated in Fig.~\ref{fig:buildingblocks}.

As discussed earlier, physics-based ASICs differ from conventional ASICs in that they relax certain requirements that normally would be expected to hold approximately, including statelessness, unidirectionality, determinism, and synchronization. Among physics-based ASICs, we may roughly categorize devices by the subset of such requirements that are relaxed.

A number of paradigms have been proposed where circuit components in ASICs are deliberately made stateful, sometimes depending on prior history over a long period. A representative example is circuits using memristors \cite{halawani2018memristor, yan2016neuromorphic}, whose resistances depend on the amount of charge that has flowed through them. Other components also may exhibit memory effects when used in analog circuits \cite{sanchez2014floating, dillavou2022selflearning, dillavou2024machine}, removing the assumption of statelessness.

Bidirectional couplings are common in ASICs implementing Ising machines (both digital and analog)~\cite{baeising, chou2019analog, lo2023ising, moy20221, Aadit_2022, wang2017oscillator}, as well as analog devices designed to solve problems in linear and nonlinear algebra and (possibly stochastic) differential equations \cite{sun2019solving,melanson2025thermodynamic, huang2017hybrid}. Mutual interactions between physical degrees of freedom are also used in nonlinear photonics-based platforms \cite{yanagimoto2025programmable} and self-adjusting resistor networks~\cite{dillavou2022selflearning,dillavou2024machine}.

As the suppression of stateful behavior and of bidirectional information flow requires dissipation, we may expect that when these requirements are relaxed, higher energy efficiency may be achieved. Taking this idea to the extreme, reversible computing attempts to reduce dissipation significantly by avoiding any erasure of information \cite{ray2023gigahertz, frank2005introduction}. It is interesting to note that quantum computing~\cite{feynman1982simulating}, as a subset of reversible computing, exhibits bidirectional flow of information between interacting qubits.

There has also been a growing interest in nondeterministic ASICs, both analog and digital~\cite{mansinghka2009natively}. In the digital case, there is a large body of work on p-bits~\cite{chowdhury2023full, Nikhar2024-dc, Camsari_2019}, which are binary variables undergoing a continuous time Markov process (CTMC). Magnetic tunnel junctions (MTJs) display bistable stochastic behavior in voltage, and may be used as sources of analog or digital randomness \cite{camsari2017implementing}. Similarly, thermodynamic computers have employed stochastic dynamics of continuous variables (i.e., Brownian motion) using analog circuits~\cite{melanson2025thermodynamic,conte2019thermodynamic,hylton2020thermodynamic,coles2023thermodynamic_published,aifer2024_TLA,duffield2025thermodynamic,aifer2024_TBI_pub,donatella2024thermodynamic,aifer2024error,donatella2025scalablethermodynamicsecondorderoptimization,lipka2024thermodynamic}.

In a number of physics-based ASIC technologies including p-bits, there is no centralized clocking, and different signals in a single device will vary asynchronously \cite{sutton2020autonomous,dillavou2024machine}. There are also ASICs that make use of polysynchronous clocking, where instead of a single centralized clock there are multiple local clocks, which are not perfectly synchronized with each other \cite{najafi2016polysynchronous}.

\subsection{Intuition for performance advantage}

As was remarked on earlier, conventional ASICs incur time and energy costs associated with guaranteeing that the requirements of statelessness, unidirectionality, determinism, and synchronization are approximately satisfied. In general, these costs are often worthwhile, as they allow for very modular design of computing systems that can be used for a variety of purposes. However, for problems of a given type, there often exist algorithms or methods of solution that do not rely on these properties. In such cases, it may be beneficial to design an ASIC to solve a problem of that specific type, and where design constraints associated with ensuring statelessness, unidirectionality, and/or determinism are relaxed.

Practically speaking, this may take the form of increasing the clock rate past the point where stateless or deterministic behavior can be relied upon. Similarly, supply voltages may be reduced, also creating non-deterministic behavior in return for lower power consumption. Indeed, it is a common feature of physics-based ASICs that they typically save on power and energy costs~\cite{stern2024training} by relaxing the aforementioned constraints.

Interestingly, we also often observe that it is possible to fuse many operations into one when the natural dynamics of the system are used in a computation. That is, we see that in some sense the physical dynamics are performing part of the computation ``automatically'' (e.g., solving linear algebra~\cite{aifer2024_TLA} or optimization~\cite{vadlamani2020physics} problems.) This provides some intuition for where time and energy savings can occur.

While there remains much work to do in scaling various approaches to physics-based ASICs, there are promising signs of the potential for significant advantages in time and energy cost.

\section{Design Strategies}

\subsection{Top-Down vs. Bottom-Up}

\begin{figure}[t]
    \centering
    \includegraphics[width=0.49\textwidth]{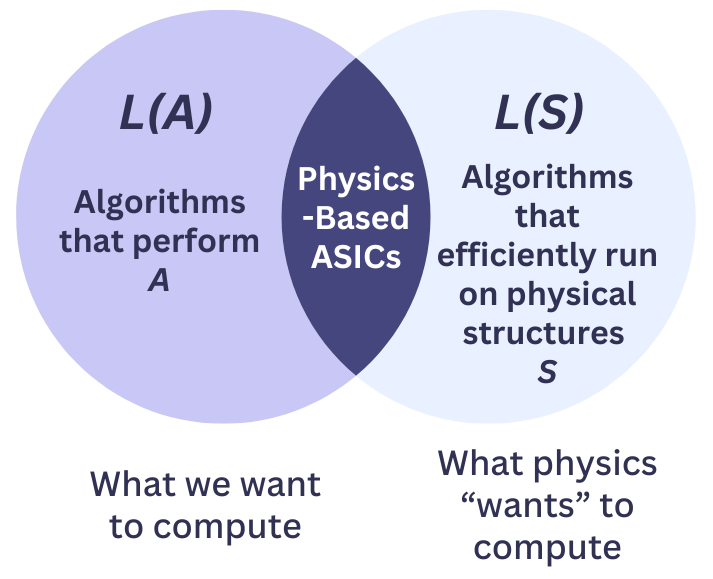}
    \caption{\textbf{Designing physics-based ASICs.} The top-down approach gives a set $L(A)$ of algorithms that can run a desired application $A$. The bottom-up approach gives a set $L(S)$ of algorithms that can efficiently run on some physical structures $S$. The basic design principle to maximize the overlap between these two sets.}
    \label{fig:VennDiagram}
\end{figure}

Designing physics-based ASICs is challenging. A principled strategy often involves considering the intersection between the top-down and bottom-up perspectives, as depicted in Fig.~\ref{fig:VennDiagram}. In the top-down approach, one starts from a key application $A$ of wide interest or great impact (e.g., generative AI for images or materials). Then one maps this application to algorithm space, i.e., mapping out the set of algorithms $L(A)$ that could potentially run that application (e.g., diffusion models, transformers, etc.). Alternatively, in the bottom-up approach, one starts from a basic physical structure $S$, like one of the structures shown in Fig.~\ref{fig:buildingblocks}. Then one determines the mathematical primitives that can be efficiently computed with those structures, $P(S)$. In turn one forms algorithms out these primitives to find the set of algorithms $L(S)$ that can be efficiently run with these structures. The goal is to maximize the overlap between the sets $L(A)$ and $L(S)$. Often this involves considering multiple different candidates structures for $S$ and then choosing the one that leads to the best overlap (in algorithm space) with a target application~$A$.

Different fields might benefit from this strategy. For example, it is common in quantum computing to primarily take the bottom-up perspective of focusing on physical structures (e.g., ions, atoms, superconducting circuits, etc.). At the same time, it is reasonable to view quantum computing as a special case of physics-based ASICs. Hence using the framework above in the quantum context could be a useful perspective for designing quantum ASICs, by keeping in mind the set of algorithms $L(A)$ that one hopes to intersect with for target application $A$.

\subsection{Performance metrics}

We can refine this strategy by making it more quantitative. Namely, we must clarify what it means for an algorithm to run ``efficiently'' on some hardware. While there are a range of potential performance metrics, two key metrics are runtime and energy consumption. For a given algorithm $\ell$, determining whether or not $\ell \in L(S)$ can involve comparing the runtime and energy consumption of $\ell$ on two hardwares: the state-of-the-art (SOTA) digital hardware (often a GPU) and the hardware built from structure $S$. For this purpose, we define the ratios:
\begin{equation}
\label{eqn:R_T}
R_T(\ell) = \frac{\text{Runtime on SOTA}}{\text{Runtime on $S$}}
\end{equation}
and
\begin{equation}
\label{eqn:R_E}
R_E(\ell) = \frac{\text{Energy consumed on SOTA}}{\text{Energy consumed on $S$}}\,.
\end{equation}
A reasonable criterion for including algorithm $\ell$ in the set $L(S)$ would be the condition that either $R_T(\ell)$ or $R_T(\ell)$ is greater than one. On the other hand, if neither of these ratios is bigger than one, then algorithm $\ell$ is not considered efficient on hardware $S$. 

A caveat to consider is that time and energy can be traded off for one another. This is why it is important to consider both ratios, because it is often possible to increase one ratio at the expense of reducing the other. To address this, one might consider a more stringent criterion that both ratios are larger than one, for $\ell$ to be considered efficient on $S$.

\subsection{Amdahl's Law}

In practice, algorithms are composed of multiple steps, and only a fraction of the computations in an algorithm can be run efficiently on the physics-based ASIC. For example, in a Kalman filter algorithm there are both matrix inversions and matrix-vector multiplications (MVMs), and one may one wish to use the ASIC for the inversions, while a GPU may be used for the MVMs. In this case, Amdahl's Law places a restriction on the performance gain one can obtain with the ASIC. Let $x$ be the fraction of the algorithm's runtime $T$ that is devoted to the computation that can be accelerated on the ASIC. Then $(1-x)T$ is the minimum runtime that can be achieved by employing the ASIC, and hence the maximum speedup is only a factor of $1/(1-x)$. A similar argument holds for the maximum gain in energy efficiency.

\subsection{Algorithmic Co-Design}

Because of Amdahl's Law, careful thought must be taken to design algorithms for a given hardware paradigm. It is interesting to take the following perspective, e.g., regarding today's SOTA algorithms for AI applications. These algorithms have implicitly been co-designed to a specific hardware platform, namely GPUs. Transformers, for example, are ideally matched to GPUs because they perform large amounts of parallelizable matrix math, and GPUs are specifically built for parallel matrix math. In this sense, GPUs have benefitted from a massive community of researchers that have co-designed algorithms for their platform. 

Likewise, physics-based ASICs will benefit from academic research on algorithmic co-design. For a given algorithmic framework, there are hyperparameters that allow one to push complexity from one subroutine to another (e.g., from sampling to optimization, or from neural network complexity to time evolution of a dynamical system). The key is to push complexity in such a way that increases the fraction $x$ appearing in Amdahl's Law. It is thus crucial to think an algorithm $\ell (h)$ as only being defined up to its hyperparameters $h$. Moreover, just because some algorithm $\ell(h)$ does not show performance advantage for hardware $S$, it does not preclude the possibility of pushing complexity around to obtain a modified algorithm $\ell(h')$ that does show advantage. It is, thus, worthwhile replacing the performance metrics in Eqs.~\eqref{eqn:R_T} and \eqref{eqn:R_E} with
$\hat{R}_T(\ell) = \max_{h} R_T(\ell(h))$ and $\hat{R}_E(\ell) = \max_{h} R_E(\ell(h))$, which represent a maximum over all efforts to co-design the algorithm $\ell$ to the hardware $S$.

\subsection{Physical machine learning}

One approach to co-designing algorithms and hardware is to machine-learn algorithms directly at the hardware level -- an approach we call ``physical machine learning'' (PML). PML typically involves a supervised learning process in which the computation performed by a given physical hardware is learned by directly optimizing the available physical parameters of the hardware (e.g., tunable conductances) such that the end-to-end transformation of data through the physical hardware best-matches a training data set. Mathematically, input data to the hardware $\vec{x}$ are encoded via some subset of programmable parameters (e.g., the voltages applied to a subpart of the hardware, such as input current sources), and, after some time, a (typically different) subset of degrees of freedom of the physical hardware are measured (e.g., the currents leaving a set of defined output wires) to produce an output vector $\vec{y}$. The process of physical machine learning then involves using an optimization algorithm to set controllable parameters, $\vec{\theta}$, of the hardware that affect the effective computation from $\vec{x}$ to $\vec{y}$, i.e., $\vec{y} = f_p(\vec{x},\vec{\theta})$, where $f_p$ represents the transformation between input and output due to the time-evolution of the hardware. For example, tunable parameters could be voltages applied to transistors between the input and output currents, which modify how currents flow through the hardware. In many forms of PML, tunable parameters may be encoded in the form of digital pre- or post-processing. For example, in physical reservoir computing \cite{tanaka2019recent,nakajima2020physical} -- a pioneering concept for PML -- the physical transformation is (in the simplest, non-recurrent case), $\vec{y} = W(\vec{\theta})f_p(\vec{x})$. Here, $W(\vec{\theta})$ is a linear matrix that is learned (usually by linear regression). By appropriately optimizing the weights of $W$, desired nonlinear functions may be approximated by a linear combination of ``features'' (i.e., functions of the input $\vec{x}$) which are naturally computed by the physical hardware (described as a ``reservoir'' in physical reservoir computing). Other forms of PML learn parameters of the physical hardware instead of, or in addition to, such digital post-processing, such as physical neural networks \cite{wright2022deep,momeni2024training,momeni2023backpropagation}, variational quantum algorithms \cite{cerezo2020variationalreview}, and ``in materio'' computing \cite{miller2014evolution,chen2020classification,ruiz2020deep}.

On one hand, PML provides a potentially elegant solution to the joint optimization of hardware and software, since algorithms are effectively learned directly from the space of computations the hardware natively provides. However, in demonstrations so far, the algorithms learned by PML have typically been quite simple, owing either to the difficulty of performing the optimization of parameters, or to limited expressive power of the physical hardware itself. Optimization in PML is difficult in part because, unlike modern artificial neural networks which have been systematically engineered to behave well when stochastic gradient descent is applied to learn their parameters (e.g., by architectural innovations like residual connections), far less engineering has been realized for physical hardware learning, and thus many hardware ansatzes (i.e., the specific form of $f_p$) exhibit more challenging optimization landscapes, exhibiting, e.g., barren plateaus~\cite{larocca2024review} that make gradient descent ineffective. Furthermore, optimizing a physical hardware directly encounters difficulties because the hardware generally differs from any idealized simulation -- this ``sim2real'' gap means that simply performing optimization with a simulation of the hardware often fails. As a consequence of these challenges, an important open challenge for the subfield of PML is to develop effective learning algorithms that can be used either within a separate processor to configure $\vec{\theta}$ efficiently, or -- ideally -- to use the physical hardware itself for this purpose, i.e., physical learning \cite{dillavou2022selflearning,dillavou2024machine,momeni2024training}. 

\subsection{Physical Learning}

The most powerful form of PML is one in which both inference (i.e., physical computation of $\vec{y} = f_p(\vec{x},\vec{\theta})$) and learning (i.e., determining the optimal choice of parameters $\vec{\theta}$) are done within the physical hardware itself. Compared to hardware that performs only inference, it is clearly more difficult to design scalable hardware (and scalable learning algorithms) in which both of these functions can be realized. But this difficulty is unquestionably worth overcoming: Solving the challenge of physics-driven learning could allow for remarkably scalable Physics-based ASICs that simply learn to perform desired computations directly, even without supervision from digital computers. This could allow compact, efficient neural network computations with vastly more trainable parameters than modern digital systems. 

For physics-driven learning in electronic hardware, it is typically necessary to have local rules~\cite{bosch2025local} to update the edges, just as neurons in the brain self-update based on local conditions without knowledge of the states of all the other neurons. One important class of such rules has been developed for electronic, fluidic, or mechanical networks that equilibrate according to an optimization principle~\cite{scellier2017equilibrium,Stern2021coupled}.  For electronic networks in the lab, circuitry have been built on each edge that implements the local learning rule to adjust its conductance~\cite{dillavou2022selflearning, dillavou2024machine}. After training is done, subsequent computation (inference) is then done physically just by supplying input voltages, letting the system equilibrate, and reading the output voltages.  Design issues then relate to hardware and choices for adjustable edges, learning circuitry, network architecture, and -importantly- how to implement at scale on-chip. Physical learning can also be accomplished in a variety of other systems~\cite{Stern2023noneuron}.

\section{Applications}

\begin{figure}[t]
    \centering
    \includegraphics[width=0.49\textwidth]{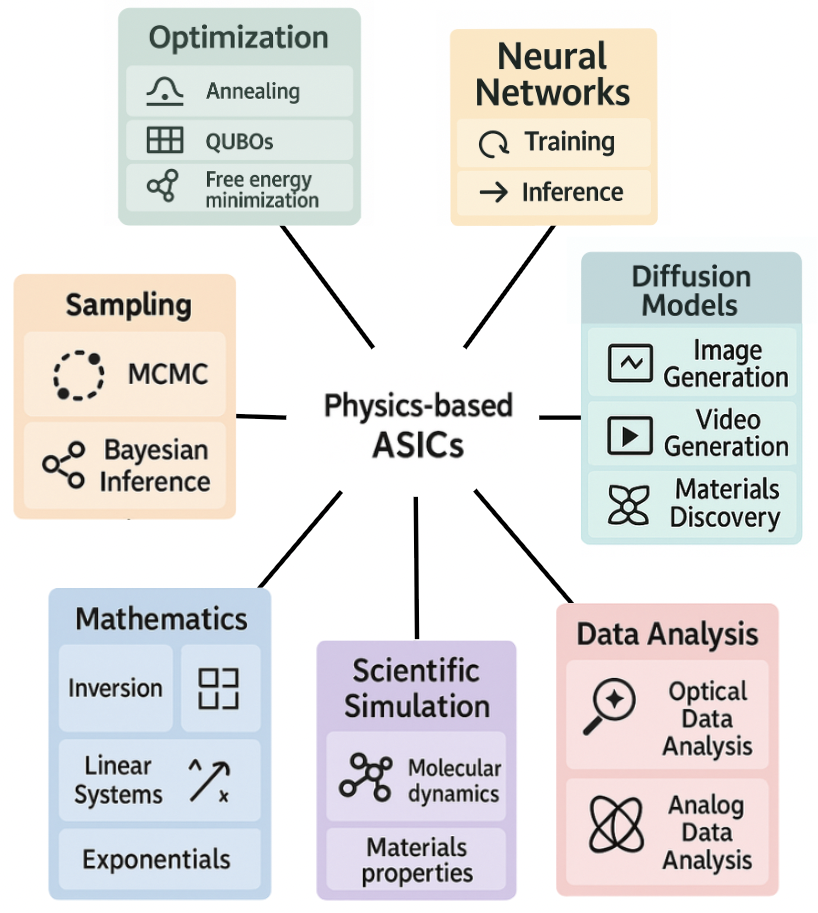}
    \caption{\textbf{Applications of Physics-based ASICs}. Some applications are physics-inspired (such as sampling and optimization). Other applications are physics-based (like scientific simulation and analog data analysis). Abstract applications in mathematics are also relevant.}
    \label{fig:applications}
\end{figure}

Figure~\ref{fig:applications} shows some applications that physics-based ASICs will impact. These devices are naturally suited to applications either inspired by or based in the physical world.

\subsection{Physics-inspired applications}

Many algorithms are physics-inspired. This is likely because the humans that have developed these algorithms have a strong intuition for physics, and also historically physics was a early application of focus. While modern applications like AI and finance are more abstract, they nevertheless often use physics-inspired algorithms in practice.

\subsubsection{Artificial Neural Networks}

The 2024 Nobel Prize in Phyics was awarded to Hopfield and Hinton for foundational work that enabled machine learning with artificial neural networks (ANNs). Although it is debatable whether ANNs are physics-inspired or biology-inspired, early ANNs like the Hopfield network and the Boltzmann machine were derived from the statistical physics of spin systems.

While ANNs used in mainstream machine learning differ greatly from the biological neural networks in brains, they nonetheless are an unusually suitable calculation for noisy analog hardware. First, modern ANNs rely heavily on a limited set of operations that are repeated many times, such as matrix-matrix and matrix-vector multiplications. This means that physics-based ASICs that accelerate just a limited class of calculations in this category may provide significant advantages for neural network inference or training \cite{chen2020survey}. Second, while modern ANNs are usually implemented in high-precision digital computers, they have nonetheless been shown to be extremely resilient to noise -- ANNs can often be trained to operate with very low (even binary) precision weights and activations with minimal performance loss \cite{qin2020binary}, and noise is often utilized in training (e.g., in the form of dropout) to improve generalization and make the learned neural network more robust to adversarial attacks \cite{poole2014analyzing}. Finally, ANNs exhibit remarkable improvements as they are afforded more compute \cite{kaplan2020scaling}, such as by increasing the number of learnable parameters (and computations per inference), training longer, or by other methods that increase compute utilization. This combination of features makes ANNs exceptionally well-suited to acceleration by special-purpose, noisy analog hardware, and their rapidly expanding applications suggests a formidable commercial incentive to do so.

\subsubsection{Diffusion Models}

One of the clearest examples of a physics-inspired algorithm is the diffusion model. The original work~\cite{pmlr-v37-sohl-dickstein15} on this topic noted the deep connection to non-equilibrium thermodynamics. As time reversals of stochastic processes are often considered in stochastic thermodynamics, it was realized that the same formalism could be used to reverse a process of adding noise to data, resulting in a generative model. Over the past few years, diffusion models have emerged as a state-of-the-art method for generative modeling of images, videos, molecular structures, and materials. The naturally occurring stochasticity in silicon makes physics-based ASICs well-suited to running diffusion models~\cite{coles2023thermodynamic_published,whitelam2025generative}. Moreover, diffusion models allow us to relax the constraints of deterministic operation to achieve more efficient performance with physics-based ASICs.

\subsubsection{Sampling}

More broadly, the problem of sampling from a desired probability distribution can be approached using physics-inspired algorithms. Physical systems like gases or magnetic spins naturally evolve towards equilibrium, where the microstates are Boltzmann distributed and the likelihood of a state is exponentially suppressed by its energy. Modern sampling methods mimic this behavior to generate samples from complex distributions in non-physical domains, such as machine learning or Bayesian inference. Algorithms like Markov Chain Monte Carlo (MCMC)~\cite{gilks1995markov} and its variants (e.g., Metropolis methods, Langevin Monte Carlo, and Hamiltonian Monte Carlo~\cite{neal2011mcmc}) emulate the random walks of particles in a physical system and can sample efficiently in high-dimensional spaces. Physics-based ASICs are poised to make both discrete- and continuous-variable sampling more efficient, effectively by returning sampling algorithms to their roots. Namely, these ASICs employ actual physics systems whose dynamics (e.g., thermodynamic relaxation) implement Langevin Monte Carlo, Bayesian inference, or other sampling protocols. Both probabilistic computing~\cite{chowdhury2023full, Nikhar2024-dc, Camsari_2019,camsari2017implementing} with Ising machines and thermodynamic computing~\cite{melanson2025thermodynamic,conte2019thermodynamic,coles2023thermodynamic_published,aifer2024_TLA,aifer2024_TBI_pub,donatella2024thermodynamic,aifer2024error} with stochastic circuits are promising approaches for sampling-based applications.

\subsubsection{Optimization}

Optimization also draws profound inspiration from physics, as physical systems naturally perform optimization~\cite{vadlamani2020physics}. Thermodynamic systems evolve toward configurations that minimize free energy, driving phase transitions like crystal formation and protein folding. This principle of free energy minimization mirrors optimization algorithms that search for the global minimum of a cost or loss function, akin to a physical system settling into its most stable state. For instance, simulated annealing explicitly emulates the cooling process of metals that eliminates atomic-level defects. Certain physics-based ASICs can perform annealing algorithms like this where an abstract loss function is encoded into a physical energy function~\cite{coles2023thermodynamic_published}. Similarly, Langevin dynamics can be viewed as a Wasserstein gradient flow. Physics-based ASICs that evolve under Langevin dynamics are performing gradient descent in Wasserstein space (i.e., a metric space on probability density functions) and hence can be used to optimize over probability distributions. Combinatorial optimization, which involves finding the optimal solution from a discrete set of possibilities, can be viewed as physics-inspired through its connection to Ising models. Namely, quadratic unconstained binary optimization (QUBO) problems can be mapped to the energy function of an Ising model. Ising machines~\cite{baeising, chou2019analog, lo2023ising, moy20221, Aadit_2022, wang2017oscillator} make use of this deep connection to efficiently solve QUBO problems, and extensions to mixed-variable optimization are possible~\cite{mourgias2023analog}. Finally, Kirchhoff's laws for electronic circuits can be interpreted as an optimization problem where the system naturally minimizes energy dissipation (subject to constraints). This has been exploited to solve optimization problems with a resistive network acting as a physics-based ASIC that both trains itself and does the desired computation~\cite{dillavou2022selflearning, dillavou2024machine}.

\subsection{Physics-based applications}

\subsubsection{Scientific simulation}

Some of the clearest applications of physics-based ASICs involve simulating the physical world. Designing novel materials and predicting their properties is an exciting task, but it is challenging for today's computing hardware. Physics-based ASICs are poised to accelerate materials discovery, both through physics-inspired generative AI (e.g., diffusion models~\cite{pmlr-v37-sohl-dickstein15}) and through enhanced characterization of material properties via dynamical simulations. Molecular Dynamics (MD) simulation methods~\cite{hollingsworth2018molecular} are already widely used in industry, e.g., in catalyst design for ammonia synthesis and environmental protection. Feynman argued that we should simulate physics with physical systems~\cite{feynman1982simulating}, and MD is good example where physics-based ASICs could accelerate the dynamical simulation of molecules and materials. This could involve accelerating primitives including Langevin dynamics~\cite{bussi2007accurate}, umbrella sampling~\cite{kastner2011umbrella}, and transition path sampling~\cite{bolhuis2002transition}. We also highlight the often overlooked regime of mesoscopic simulation, where quantum effects are washed out, and classical stochastic thermodynamics is the appropriate framework. Physics-based ASICs will play a key role in mesoscopic simulation (e.g., for self-assembly in nanostructured materials and rheology of non-Newtonian fluids), likely as a component of multiscale modeling of complex engineering processes. At a deeper level, there are unsolved scientific questions at the mesoscopic level, such as the origin of life. Physics-based ASICs could test, e.g., England's theory of dissipation-driven adaptation~\cite{england2015dissipative} and the physics-based emergence of self-replication~\cite{england2013statistical}, explaining how life emerged on Earth.

\subsubsection{Analog Data Analysis}

Physics-based neural networks have shown promise for analyzing data that are inherently analog in nature~\cite{wright2022deep}. For example, optical data is naturally analyzed by optical neural networks~\cite{Wang2022-nr}, and analogous statements hold for audio data or analog electrical data~\cite{dillavou2022selflearning, dillavou2024machine}. Physics-based neural networks avoid the overhead of having to transduce analog signals into the digital domain by performing the analysis directly in the analog domain. These applications will become especially important as AI becomes more multimodal and more embodied in the physical world (e.g., via robotics), as highlighted by Jensen Huang~\cite{huang2025physicalai}.

\section{Roadmap and Challenges}

We expect the adoption of physics-based ASICs to follow three phases. In the first phase, research groups will leverage proof-of-concept hardware to show that their physics-based ASIC architectures have the potential to achieve greater performance than state-of-the-art methods run on CPUs and GPUs. Next, key scalability problems will need to be resolved to enable physics-based ASICs to solve problems of similar scale and complexity to existing hardware solutions. Finally, these scaled-up physics-based ASICs need to be integrated into systems, with software abstractions designed to make them easy to use to run key computational workloads.

\subsection*{Phase 1: Demonstrate domain-specific advantage}

The most important driver of adoption for physics-based ASICs will be their performance and energy efficiency when running key computational workloads. Hence, one of the first goals of any physics-based ASIC project should be to demonstrate a path to running one of those workloads better than conventional CPU- or GPU-based methods.

\subsubsection*{Speed-up for key applications}

For some problems, relatively small-scale prototypes of physics-based ASICs can demonstrate better performance than CPU- or GPU-based solvers. For example, latch-based Ising machines were able to minimize Ising Hamiltonians over 1000 times faster than CPU-based solvers for problems with 1440 Ising spins~\cite{baeising}. However, for larger problems, these prototypes often cannot achieve the same speed-up, due to the cost of loading data onto and reading data off of the physics-based ASIC. This highlights memory bandwidth and scalability as key limitations of prototype systems.

Another route to prove a potential speed-up is by demonstrating a key scaling advantage. For example, analog Ising machines based on coupled oscillators were predicted to outperform a GPU-based solver at a scale of about 150 spins or larger~\cite{chou2019analog}. Similarly, thermodynamic computing achieved a better asymptotic complexity for linear algebra~\cite{aifer2024_TLA,duffield2025thermodynamic} and Bayesian inference~\cite{aifer2024_TBI_pub} tasks compared to state-of-the-art digital methods, and these complexity advantages carry over to higher-level applications such as neural network training~\cite{donatella2024thermodynamic,donatella2025scalablethermodynamicsecondorderoptimization}, which is extremely computationally expensive on GPU. 

Nevertheless, Moore’s Law progress in the past has largely been driven by shrinking scaling prefactors even when the asymptotic scaling has stayed fixed. Hence traditional engineering innovation aimed at prefactors remains important.

\subsubsection*{Energy efficiency}

Physics-based ASICs also have the potential to achieve far greater energy efficiency than GPU-based solvers due to the more natural mapping of certain applications to physics-based hardware. Optical neural networks have been shown to perform classification tasks with less than one detected photon per scalar multiplication~\cite{Wang2022-nr}. This results in a fundamental energy advantage over conventional approaches based on digital electrical circuits. Similarly, a coupled-oscillator-based Ising machine with all-to-all connections was able to solve combinatorial optimization problems with 1-2 orders of magnitude lower energy consumption than state-of-the-art algorithms running on CPUs~\cite{cilasun2025coupled}. In addition, physical computation in analog electronic networks of self-adjusting resistors potentially offers a million-fold energy savings compared to digital computation~\cite{dillavou2022selflearning, dillavou2024machine}.

\subsection*{Phase 2: Architect scalable physical substrates}

The majority of physics-based ASICs demonstrated in the literature are relatively small-scale, especially compared to their conventional digital counterparts. This work is valuable in demonstrating the viability of physics-based ASICs as a concept, but additional work is required to scale these designs to the point where they can tackle real-world problems. The following discussion gives a few potential strategies for scalability, although not an exhaustive list of possible approaches.

\subsubsection*{Tile-based ASIC design}

For example, ``Field-Programmable Ising Arrays'' were proposed~\cite{hutchinson2024fpiafieldprogrammableisingarrays}, and they leverage a tile-based hierarchy to achieve greater efficiency and reconfiguration. Each tile consists of dense all-to-all analog coupling, while inter-tile connections are all-digital and sparse. This limits the size of the analog coupling circuits, preventing circuit parasitic effects and noise from significantly degrading their performance.

Tile-based architectures and other hierarchical architectures also have practical advantages in terms of physical implementation. The tile design can be designed, synthesized, and routed independently of other tiles, and then placed into a network-on-chip (NoC) to communicate with other tiles. This is especially valuable when each tile includes analog or mixed-signal components; designing and simulating a mixed-signal chip on the scale and complexity of a GPU without a hierarchical architecture would be extremely difficult.

\subsubsection*{Reconfigurable interaction terms}

Many physics-based ASICs also struggle to support dense problems. Often, they will only support a fixed graph topology and require software to map the arbitrary problems to that fixed graph topology. This process, known as Minor Embedding \cite{choi2008minorembeddingadiabaticquantumcomputation}, is computationally expensive and may fail for large or difficult problems~\cite{Aadit_2022}. Clearly, better hardware support is needed for graphs with different sparsity patterns.

Along these lines, a p-bit computing architecture was proposed~\cite{Nikhar2024-dc} that uses a reconfigurable master graph to support graphs with different sparsity patterns. This method preserves the scaling advantages of sparse connectivity scaling; the number of neighbors for each p-bit stays constant, so the hardware utilization scales linearly, and the maximum frequency stays approximately constant as the number of p-bits increases.

By leveraging modular, tile-able compute units and reconfigurable coupling, physics-based ASICs could potentially get as large as GPUs, while being able to support problems with many different sparsity patterns.

\subsection*{Phase 3: Integration into hybrid systems}

Once large-scale physics-based ASICs have been demonstrated, they need to be integrated into practical, large-scale systems, from both a hardware and software perspective.

\subsubsection*{Heterogeneous hardware platforms}

Because physics-based ASICs are specialized for solving certain kinds of computational problems, we expect them to be deployed in so-called \textit{heterogeneous systems}, which feature conventional GPUs and CPUs alongside physics-based ASICs. For example, a supercomputing system was proposed~\cite{mohseni2025buildquantumsupercomputerscaling} that leverages multiple networked probabilistic processors alongside conventional GPUs and quantum processors. Such a system would be able to accelerate energy-based models (EBMs) efficiently, using the GPU for large matrix-vector operations like embedding calculation and gradient computation, while using the probabilistic processors for simulating stochastic neuron operation.

One major challenge engineers will face when designing heterogeneous hardware platforms with physics-based ASICs is inter-chip networking. For large problems, single blocks of computation may need to be distributed across multiple physics-based ASICs~\cite{mohseni2025buildquantumsupercomputerscaling}. Because many physics-based ASICs leverage unconventional data representations and communication methods, specialized chip-to-chip interfaces may be necessary to enable workloads to be mapped to multiple physics-based ASICs efficiently. At the same time, physics-based ASICs need to fit into heterogeneous systems that use standard interfaces, like PCIe, Ethernet, and Infiniband.

\subsubsection*{Standard software abstractions}

For physics-based ASICs to achieve wide adoption, they need to be easy to use for software engineers who may not be familiar with the underlying physical processes powering the computation. In practice, this means that physics-based ASICs should leverage standard software abstractions that users are already comfortable with, like PyTorch and JAX. For example, Python-based programming models have been developed for electrical~\cite{BelatecheDIMPLE,BelatecheSB} and photonic~\cite{Arai2020,Chen_cim-optimizer_a_simulator_2022} Ising machines.

To enable users to run more complex workloads on physics-based ASICs, a compiler layer will likely be necessary. (For example, D-Wave's Advantage has a mature software stack that can serve as inspiration for the software necessary to commercialize a physics-based ASIC~\cite{dwaveocean}.) Such a compiler will allow users to define workloads at a high level, and automatically perform the task of mapping parts of that workload to appropriate physics-based ASICs in a way that results in maximal speedup. One can envision this as a `CUDA moment' for physics-based ASICs, which requires EDA-like flows that combine front-ends for PyTorch or JAX, analog or mixed signal place-and-route, and open source simulators for design-space exploration. These tools were essential for scaling conventional electronics and will be similarly critical here. Ultimately, our vision would allow for a user to write a single program in PyTorch or JAX and have that workload automatically be compiled and run on a hybrid system consisting of CPUs, GPUs, and physics-based ASICs such that each kind of chip is used for the parts of the workload where it offers the best speed or efficiency.

\section{Conclusion}

\subsection{Vision for the field}

As conventional scaling plateaus, physics-based ASICs provide not only a viable alternative but a necessary evolution in how we compute. This new field exploits, rather than fights, natural physical processes. Standard computing spends massive amounts of energy just to satisfy assumptions that allow one to abstract away the underlying physics of silicon. By relaxing these assumptions, our field aims to address the unsustainable energy usage of today's computing hardware. Moreover, we aim to accelerate key applications that bottleneck AI workloads: sampling, generative AI, optimization, neural network training and inference, and even simulating other physical systems.

The path forward is not defined by a single architecture or a universal solution. One can envision a future where high-performance computing (HPC) platforms are composed of multiple physics-based ASICs, each specialized for different roles. For example, an HPC platform for multi-scale physical modeling can incorporate multiple ASICs operating at distinct length scales -- atomic, micro, meso, and macro -- where each scale requires hardware tuned to its specific physical regime. Similarly, an HPC platform for model-based reinforcement learning could be composed of a heterogeneous compute stack optimized for application-specific performance, with different ASICs solving different subroutines like sampling, optimization, and physical simulation.

While energy and time play fundamental roles in physics, they are also fundamental to computation. This is not a coincidence, as physics provides a unifying framework for assessing computational performance~\cite{wolpert2024stochastic}.  For example, energy-time tradeoffs (or more generally, energy-time-accuracy tradeoffs) naturally arise for physics-based ASICs~\cite{aifer2024_TLA,stern2024training}, yet such tradeoffs are also relevant for understanding the complexity of standard digital computation~\cite{valiant2023matrix}. This suggests the potential for unified treatments of computational complexity, which will be a key to comparing different paradigms on equal footing.

Physics-based ASICs offer a path for computation beyond the limits of conventional scaling. Within the next two years, we expect to see an increasing number of demonstrations of performance advantage of physics-based ASICs over standard hardware. The first demonstrations may show large energy efficiency gains for the same performance quality, while later demonstrations will likely unlock new features that otherwise would not be possible with standard hardware. The latter may include scalable approximation-free Bayesian inference for reliable AI predictions, molecular dynamics simulations at large scale with high accuracy, or rapid analysis of analog physical data processed in edge devices.

\subsection{Calls to action}

Realizing this vision will benefit from an active, growing community of enthusiastic developers. We highlight several key areas where action is needed:
\begin{itemize}
    \item Identifying a set of applications for which GPUs are not well optimized. GPUs are well-suited to parallel, but not necessarily sequential, computations. For example, applications involving simulating physical dynamics are sequential in time and likely are challenging for GPUs. While many physical simulations may be parallelized across subsystems (e.g., individual particles), there is a limit to the performance advantage that can be achieved through paralellism alone; the total runtime will always be bounded below by the length of the longest chain of sequential steps, multiplied by the latency of a single step. For this reason, ultimately performance will be limited by the latency of simulation steps rather than the total computational throughput in a highly parallel scenario.
    \item Co-designing algorithms for physics-based ASICs. Transformers have been co-designed to GPUs. Our field similarly needs a strong effort from the algorithms and applications community to create new algorithms, and push complexity around in old algorithms, to increase the performance gains for physics-based ASICs.
    \item Developing the full stack for physics-based ASICs. This new hardware will require compilers and user interfaces to obtain widespread adoption by community. Open-source software will likely be a key ingredient for adoption. Moreover, the development of simulators for physics-based ASICs will help to democratize the field.
    \item Explaining the goals and approaches of our work in a way that does not require a background in physics or electrical engineering to understand, so that we can lower the barrier to having meaningful interactions with computer scientists. 
\end{itemize}

\subsection{Urgency of the field}

As a final remark, we believe the compute-related crises society faces provide urgency to our field, from the AI energy crisis to the compute cost crisis to the end of key scaling laws. The fact that multiple crises could be addressed through a single technology presents a unique opportunity. Moreover, this technology is further motivated by the unusually fast rise of AI. It is clear that AI has been the economic driving force responsible for this new field of physics-based ASICs, and hence the two fields are intimately connected. As AI continues to move closer to the physical realm, there is the enticing prospect that physics-based ASICs will provide the physical embodiment for AI in the future.

\bibliography{bibfile.bib}

\end{document}